\documentclass{ifacconf}

\usepackage{graphicx}      
\usepackage{natbib}        

\usepackage{bm}		       
\usepackage{amsmath}       
\usepackage{amsfonts}      
\usepackage{color}		        
\usepackage{mathtools, nccmath}
\usepackage{siunitx}

\usepackage{tablefootnote} 

\usepackage{comment}

\usepackage[compatibility=false]{caption}
\usepackage{threeparttable} 
\usepackage{booktabs} 

\graphicspath{{Figures/}}

\begin{document}
\begin{frontmatter}

\title{Deep reinforcement learning-based spacecraft attitude control with pointing keep-out constraint}

\thanks[footnoteinfo]{\copyright 2025 the authors. This work has been accepted to IFAC for publication under a Creative Commons Licence CC-BY-NC-ND.}

\author[First]{Juntang Yang} 
\author[First]{Mohamed Khalil Ben-Larbi} 

\address[First]{University of Würzburg, 97074 Würzburg, Germany (e-mail: juntang.yang@ uni-wuerzburg.de, khalil.ben-larbi@uni-wuerzburg.de)}

\begin{abstract}                
This paper implements deep reinforcement learning (DRL) for spacecraft reorientation control with a single pointing keep-out zone. The Soft Actor-Critic (SAC) algorithm is adopted to handle continuous state and action space. A new state representation is designed to explicitly include a compact representation of the attitude constraint zone. The reward function is formulated to achieve the control objective while enforcing the attitude constraint. A curriculum learning approach is used for the agent training. Simulation results demonstrate the effectiveness of the proposed DRL-based method for spacecraft pointing-constrained attitude control.
\end{abstract}

\begin{keyword}
deep reinforcement learning, spacecraft attitude control, pointing constraint
\end{keyword}
\end{frontmatter}

\section{Introduction}
During spacecraft reorientation maneuvers certain sensitive instruments onboard (e.g., optical cameras) need to avoid pointing towards bright celestial objects such as the Sun. Spacecraft attitude maneuver under pointing constraints is challenging.  Approaches, such as attitude planning~(\cite{feron2001randomized, kjellberg2013discretized}) and artificial potential field (APF)-based methods~(\cite{lee2014feedback, yang2021potential}), have been proposed to address this problem, but both have key limitations. Attitude planning is computationally intensive, making it difficult for real-time onboard implementation. APF-based control methods, while efficient for onboard implementation, usually suffer from local minima which cause the spacecraft to converge to an undesired attitude. 

Deep reinforcement learning (DRL) has emerged as a promising approach for the constrained attitude control problem due to their ability to handle complex tasks effectively. Although training the agent is computationally intensive, the trained model enables rapid inference -- generating command torques in real time based on spacecraft state observations -- making it feasible for onboard implementation. While DRL has been applied to spacecraft attitude control without pointing constraints~(\cite{elkins2022bridging, gao2020satellite, Djebko2023learning, oghim2025deep}), its extension to pointing-constrained scenarios remain relatively unexplored, with only a few studies addressing this challenge~(\cite{jiang2023spacecraft,cai2024reinforcement}). 

\cite{jiang2023spacecraft} employ Deep Q-Networks (DQN) for attitude maneuver planning under forbidden and boundary constraints. Since DQN operates in discrete action spaces, the authors discretize the action space using minimum-angle slew path method. The state representation includes angular parameters that encode the information of forbidden constraints and the target attitude relative to the desired orientation. However, DQN's discrete action formulation may limit control precision in continuous dynamics. 

\cite{cai2024reinforcement} investigate satellite formation attitude control under multiple constraints (keep-out zones for optical cameras and keep-in zones for inter-satellite communication) using the Deep Deterministic Policy Gradient (DDPG) algorithm, which is suitable for continuous state and action spaces. While their simulations validate the approach, the state space design omits explicit constraint zone information, relying only on attitude error and angular velocity. This simplification restricts adaptability to varying constraint zone configurations.

Based on the above observation, this paper implements DRL for spacecraft reorientation control with a single pointing keep-out constraint, adopting the Soft Actor-Critic (SAC)~(\cite{haarnoja2018soft}) algorithm to handle continuous state and action space. The state space is designed to explicitly include a compact, physically interpretable representation of the attitude constraint zone. A tailored reward function is formulated to enforce the constraint during agent training, promoting constraint-compliant behavior. To enhance learning efficiency, a curriculum learning strategy is employed, progressively increasing task complexity from simplified to full-scale scenarios.

The rest of this paper is organized as follows: Section 2 provides details of the methodology. Section 3 presents the main simulation results and Section 4 concludes this paper.

\section{Methodology}
\subsection{Rotational kinematics and dynamics}
In this paper, unit quaternions are used as the attitude parameterization. The rotational kinematic and dynamic equations are given as follows~(\cite{markley2014fundamentals})
\begin{equation}
\label{eq:kinematics_q}
    \dot{\bm{q}} = \frac{1}{2}\bm{q}\otimes\bm{\omega}
\end{equation}

\begin{equation}
\label{eq:dynamics}
    I\dot{\bar{\omega}} =  - \bar{\omega}^{\times} I \bar{\omega} + \bar{\tau}
\end{equation}
where $\bm{q} = \left(q_0, \bar{q}\right)$ is the spacecraft attitude expressed in unit quaternion with $q_0 \in \mathbb{R}$ and $\bar{q} = [q_1, q_2, q_3]^T \in \mathbb{R}^{3}$ as the scalar part and vector part, respectively, $\bm{\omega} = \left(0, \bar{\omega}\right)$ is the vector quaternion form of the angular velocity (expressed in the body frame) $\bar{\omega} = [\omega_1, \omega_2, \omega_3]^T\in \mathbb{R}^{3}$,  \(I \in \mathbb{R}^{3 \times 3}\) is the moment of inertia of the spacecraft, \(\bar{\tau} = [\tau_1, \tau_2, \tau_3]^T \in \mathbb{R}^{3}\)  is the control torque expressed in the body frame, \(\bar{\omega}^{\times} \in \mathbb{R}^{3 \times 3}\) is defined as
\begin{equation}
\bar{\omega}^{\times}  = 
\left[\begin{matrix}
0 & -\omega_3 & \omega_2 \\
\omega_3 & 0 & -\omega_1 \\
-\omega_2 & \omega_1 & 0
\end{matrix}\right]
\end{equation}
and $\otimes$ is quaternion multiplication (see Table~\ref{tab:quaternion_op} where \(\bm{a} = \left(a_0,~\bar{a}\right)\) and \(\bm{b} = \left(b_0,~\bar{b}\right)\)).

\begin{table}[h]
\caption{\label{tab:quaternion_op} Quaternion operations}
\centering
\begin{tabular}{lc}
Operation & Definition \\
\hline
Multiplication & $\bm{a}\otimes\bm{b} = \left(a_0 b_0-\bar{a}\cdot\bar{b},~a_0 \bar{b} + b_0 \bar{a} + \bar{a} \times \bar{b}\right)$ \\
Conjugate &  $\bm{a}^* = \left(a_0,~-\bar{a}\right)$ \\
\hline
\end{tabular}
\end{table}

\subsection{Pointing keep-out zone}
\begin{figure}[h]
\centering
\includegraphics[width=.4\textwidth]
{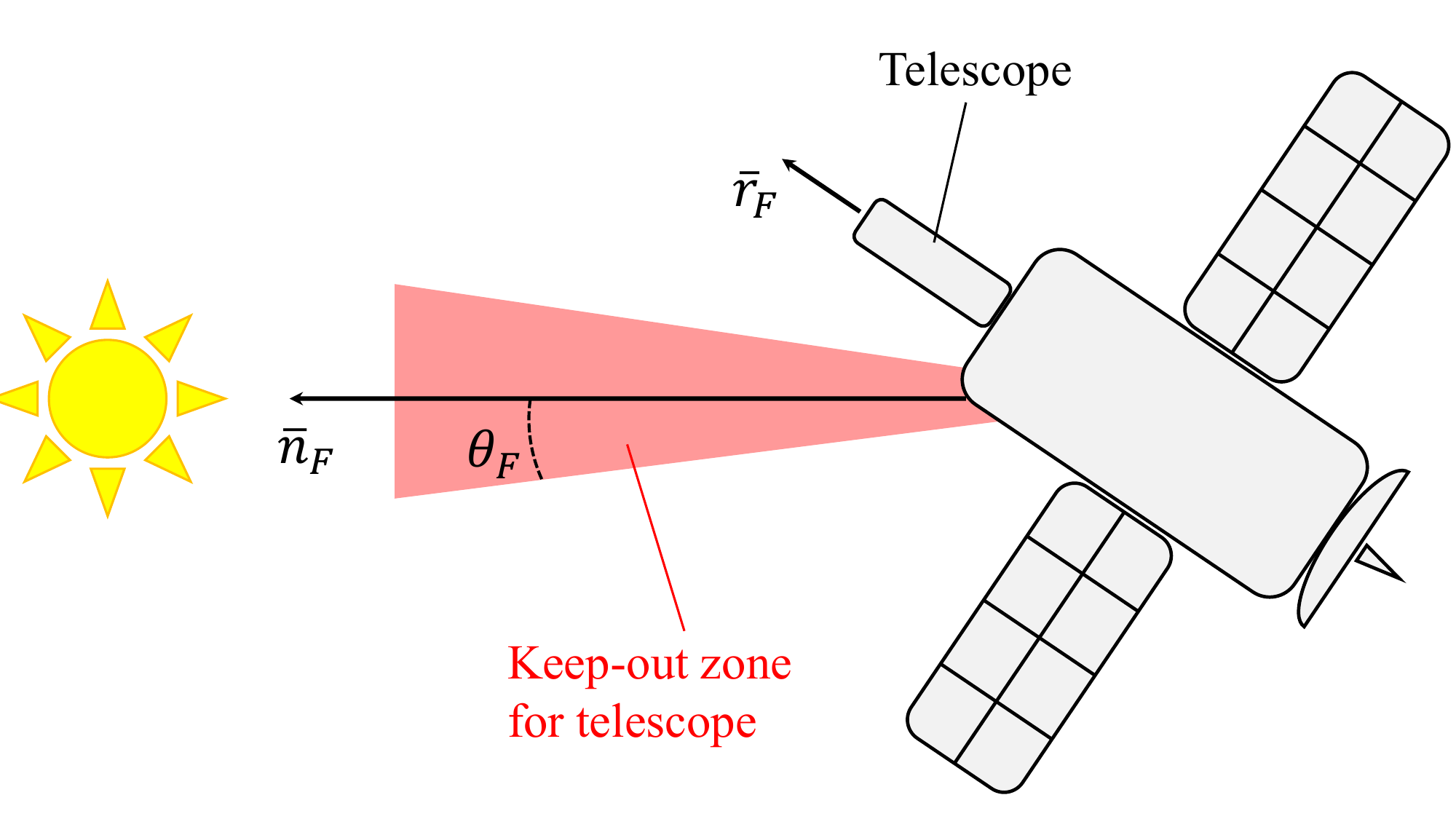}
\caption{Spacecraft sketch with keep-out zone for telescope.}
\label{fig:attitude_constraint_zones}
\end{figure}

A pointing keep-out zone or forbidden zone (as illustrated in Figure~\ref{fig:attitude_constraint_zones}) is certain inertial direction which the optical sensitive instrument (e.g., a telescope) onboard the spacecraft must avoid pointing towards during attitude maneuvers.

In Fig.~\ref{fig:attitude_constraint_zones}, the unit boresight vector of the telescope, $\bar{r}_F$, is required to stay outside the red shaded attitude forbidden zone characterized by $\bar{n}_F$ and $\theta_F$. This requirement can be mathematically formulated as
\begin{equation}
	\theta >  \theta_F
\end{equation}
where $\theta = \arccos(\bar{r}_F \cdot \bar{n}_F)$ is the angle between the unit boresight vector and the unit vector representing the direction to avoid. Alternatively, the requirement can also be formulated as
\begin{equation}
	\label{eq:requirement_f}
	\bar{r}_F \cdot \bar{n}_F - \cos \theta_F < 0
\end{equation}

\subsection{Deep reinforcement learning}
Reinforcement learning (RL) is a paradigm within machine learning in which an agent learns to make decisions by interacting with its environment to maximize cumulative rewards~(\cite{sutton2018reinforcement, arulkumaran2017deep}). 

When the Markov property is satisfied, a RL problem can be described in the framework of Markov decision process (MDP), formally defined by the tuple~$\left(S,A,R,P,\gamma\right)$. In this formulation, $S$ represents the state space, $A$ denotes the action space, $R: S \times A \rightarrow \mathbb{R}$ is the reward function, $P$ describes the state transition function, and $\gamma \in [0,1)$ is the discount factor. At each time step, the agent in state $s \in S$ takes an action $a \in A$. The environment responds to the agent's action, transitioning to a new state $s^{'}$, and provides a reward $R(s,a)$ as feedback. The fundamental objective in RL is to learn a policy that maximizes the expected cumulative discounted return over the decision horizon.

The detailed framework of RL for spacecraft reorientation control with a single pointing keep-out zone is described as follows.

State space: 
\begin{equation}
\label{eq:state_space}
    s_t = \left[\bm{q}_{e,t}, \bar{\omega}_t, \bar{r}^B_{F,t}, \theta_{\text{margin,t}}, \theta_t, \Delta \bar{n}^B_{F,t}, q_{e0, t-1}\right]
\end{equation}
where the subscript $t$ indicates the data at the current time step. For the sake of brevity, the subscripts $t$ is omitted in the following description of the state space.  
$\bm{q}_e =\bm{q}_d^* \otimes \bm{q}$ is the relative attitude with respect to (w.r.t.) the desired one, $\bar{\omega}$ is the angular rate, $\bar{r}^B_F$ is the unit boresight vector of the payload expressed in the body frame, $\theta_{\text{margin}} = \theta - \theta_F$ is the margin angle before the boresight vector enters the pointing keep-out-zone, $\theta$ is the angle between the boresight vector and the vector direction to avoid, and $\Delta \bar{n}^B_F = \frac{\bar{n}^B_F - \bar{r}^B_F}{\|\bar{n}^B_F - \bar{r}^B_F\|}$ is a unit vector indicating the relative direction of the avoid vector w.r.t. the boresight vector in the body frame. $q_{e0, t-1}$ is the scalar element of the unit quaternion at the previous time step. 

When $\theta_{\text{margin}} > 0$, the boresight vector is outside the keep-out zone.  When $\theta_{\text{margin}} = 0$, the boresight vector reaches the boundary of the keep-out zone. When $\theta_{\text{margin}} < 0$, the boresight vector enters the keep-out zone and violates the constraint. To avoid violation of keep-out zone constraints, it is required that $\theta_{\text{margin}} > 0$. $\Delta \bar{n}^B_F$ provides the agent the information of the relative geometry between the boresight vector and the avoid vector and enables the agent to generate the control torque in the correct direction to avoid the keep-out zone. It is expressed in the body frame since the torque in the dynamics equation is expressed in the body frame. $\Delta \bar{n}^B_F$ and $\theta$ provide the spatial information of the vector direction to avoid.

Action space: $\bar{\tau} = \left[\tau_1, \tau_2, \tau_3\right]$.  The action is scaled to the range $[-1, 1]$ during agent trainings.
 
Reward function: Inspired by the reward function design by~\cite{elkins2022bridging}, the reward function $r_t$ at the current time step is defined as follows 
\begin{equation}
\label{eq:reward_r}
r_t = 
\begin{cases}
r_{1,t} + 9,	& \phi_t \leq 0.25^{\circ}\\
r_{1,t}, 	& \text{otherwise}
\end{cases}
\end{equation}

\begin{equation}
    \label{eq:reward_r1}
r_{1,t} = 
    \begin{cases}
    \text{exp}\left(\frac{-\phi_t}{0.14*2\pi}\right) - 0.05\frac{\|\bar{\tau}_t\|}{\|\bar{\tau}_{\text{max}}\|} &\\
    \quad- 0.005\|\Delta \bar{\tau}_t\| - \text{P}_{\text{f-zone,t}},	& q_{e0, t} > q_{e0, t-1}\\ \\
    \text{exp}\left(\frac{-\phi_t}{0.14*2\pi}\right) - 0.05\frac{\|\bar{\tau}_t\|}{\|\bar{\tau}_{\text{max}}\|} &\\
    \quad- 0.005\|\Delta \bar{\tau}_t\| - \text{P}_{\text{f-zone, t}} -1, 	& \text{otherwise}
    \end{cases}
\end{equation}

where $\phi_t = \arccos (q_{e0,t})$ indicating the deviation angle between the current orientation and the desired one, $\bar{\tau}_{\text{max}}$ is the max control torque, $\Delta \bar{\tau_t} = \bar{\tau}_t - \bar{\tau}_{t-1}$, and the F-zone penalty term, $P_{\text{f-zone,t}}$, is designed as 
\begin{equation}
\text{P}_{\text{f-zone,t}} = 
\begin{cases}
\beta,	& \theta_{\text{margin,t}} \leq 0\\
\beta \exp{\left(-\alpha \theta_\text{margin,t} \right)}, 	& \text{otherwise}
\end{cases}
\end{equation}
with $\beta$ and $\alpha$ being positive values and $\theta_\text{margin}$ expressed in radians.

As shown in Eq.~(\ref{eq:reward_r}), when the attitude reaches the desired accuracy (set as $0.25^\circ$ in this paper), an extra reward of 9 is given to the agent. In Eq.~(\ref{eq:reward_r1}), the term $\text{exp}\left(\frac{-\phi_t}{0.14*2\pi}\right)$ represents the reward depending on the attitude, the term $- 0.05\frac{\|\bar{\tau}_t\|}{\|\bar{\tau}_{\text{max}}\|}$ considers the control effort, the term $- 0.005\|\Delta \bar{\tau}_t\|$ is to handle the high-frequent change of control torques, and the last term $- \text{P}_{\text{f-zone,t}}$ is to consider the pointing constraint.

State transition: The state transition function is determined by the rotational kinematics and dynamics presented in Eqs.~(\ref{eq:kinematics_q}) and (\ref{eq:dynamics}). 
 
\subsection{Soft actor-critic algorithm}
This work uses the Soft Actor-Critic (SAC) algorithm (\cite{haarnoja2018soft}) since it handles continuous state and action spaces and performs well in exploration-heavy tasks. As an off-policy actor-critic method, SAC demonstrates high sample efficiency. The algorithm optimizes a maximum entropy objective, where the policy (actor) simultaneously maximizes both the expected return and action entropy. This means that the actor aims to achieve the task and acts as randomly as possible at the same time, thereby enhancing  exploration during training.

\subsection{Agent training}

In this paper, the SAC algorithm is implemented for agent training based on the Stable-Baselines3~(\cite{raffin2021stable}). 

A time step of~\SI{0.1}{s} is used for the simulation environment and the duration per episode is set as~\SI{100}{s}. As regulation cases are considered, the initial angular rate is set as zero with small random noises during the training. The moment of inertia of the spacecraft is fixed during the training and is set as 
\[I = \left[\begin{matrix}
60 & 5 & 1 \\
5 & 50 & 2 \\
1 & 2 & 70
\end{matrix}\right] \text{kg}\cdot\text{m}^2\] 
In this paper, the boresight vector is fixed as $[1,0,0]^T$ in the body frame and the control torque components are set as $|\tau_i| \leq \SI{2}{Nm}~(i=1,2,3)$. Parameters for F-zone penalty term are set as $\beta = 10$, $\alpha = 66$. 

The agent training was performed in two phases based on a curriculum learning approach. 

In Phase 1, the agent was trained without considering the F-zone in the reward function. The initial attitude is set randomly based on the initial deviation from the desired one with the maximum initial deviation rotation angles increasing gradually from $25^{\circ}$ to $180^{\circ}$. Corresponding intermediate agents and replay buffers are saved for further training.  

In Phase 2, the best trained agent from Phase 1 was used for further training with the consideration of the F-zone in the reward function. One F-zone is set with its avoid vector $\bar{n}_F$ in the middle of the shortest rotation path from the random initial attitude (deviation angle randomly set in the range $[80^{\circ}, 180^{\circ}]$) to the desired one. The half angle of the cone $\theta_F$ is randomly set as a value between $15^{\circ}$ and $30^{\circ}$. Since the agent well trained in Phase 1 generates rotation close to the shortest rotation, the above setting of the F-zone can ensure that the keep-out zone will influence the training of the agent in Phase 2. During the training in Phase 2, the maximum deviation angle gradually increases from $80^{\circ}$ to $180^{\circ}$.

The agent training was carried out on a desktop computer with an Intel i7-14700 KF CPU, 32 GB of RAM and an NVIDIA RTX 4070 GPU. Hyperparameters of SAC for agent training are listed in Table~\ref{tab:hyperparameters}. 

\begin{table}[htbp]
\caption{\label{tab:hyperparameters} SAC hyperparameters for agent training}
\centering
\begin{tabular}{cc}
\hline
\hline
Parameter & Value \\
\hline
Batch size  & 256 \\ 
Buffer size & $10^6$ \\
Discount factor ($\gamma$) & 0.99 \\
Entropy coefficient & Auto \\ 
Learning rate (without F-zone) & 0.0001 \\
Learning rate (with F-zone) & 0.0001\\
Soft update coefficient ($\tau$) & 0.005 \\
\hline
\hline
\end{tabular}
\end{table}

The default policy of SAC in SB3 is used which is MlpPolicy using ReLU activation. The default network architecture is used which is a two layers fully-connected net with 256 units for each layer. 

\section{Simulation results}
In this section the performance of agent trained using SAC will be evaluated. 

First of all, a reorientation scenario with an initial deviation angle as $100^{\circ}$ and one F-zone is tested with two agents, one from Phase 1 (without considering F-zone) and the other from Phase 2 (with the consideration of one F-zone). Parameters for the test scenario are listed in Table~\ref{tab:data_simulation}.
\begin{table}[htbp]
\caption{\label{tab:data_simulation} Parameters for simulation}
\centering
\begin{tabular}{cc}
\hline
\hline
Parameter & Value \\
\hline
Target attitude  & (1,0,0,0) \\ 
Initial relative attitude ($\bm{q}_e$)  & (0.6428, 0.3138, -0.5892, 0.3757) \\ 
Initial angular rate ($\bar{\omega}$) & $[-5.7,-1.1,-9.9]^T*10^{-4}~\text{deg/s}$ \\
Avoid vector ($\bar{n}^I_{F}$) & $[0.703, 0.263, 0.661]^T$ \\
Half angle ($\theta_F$) & 15.20 deg \\ 
Boresight vector ($\bar{r}^B_{F}$) & $[1,0,0]^T$ \\
\hline
\hline
\end{tabular}
\end{table}

\begin{figure}[h]
\centering
\includegraphics[width=.4\textwidth]
{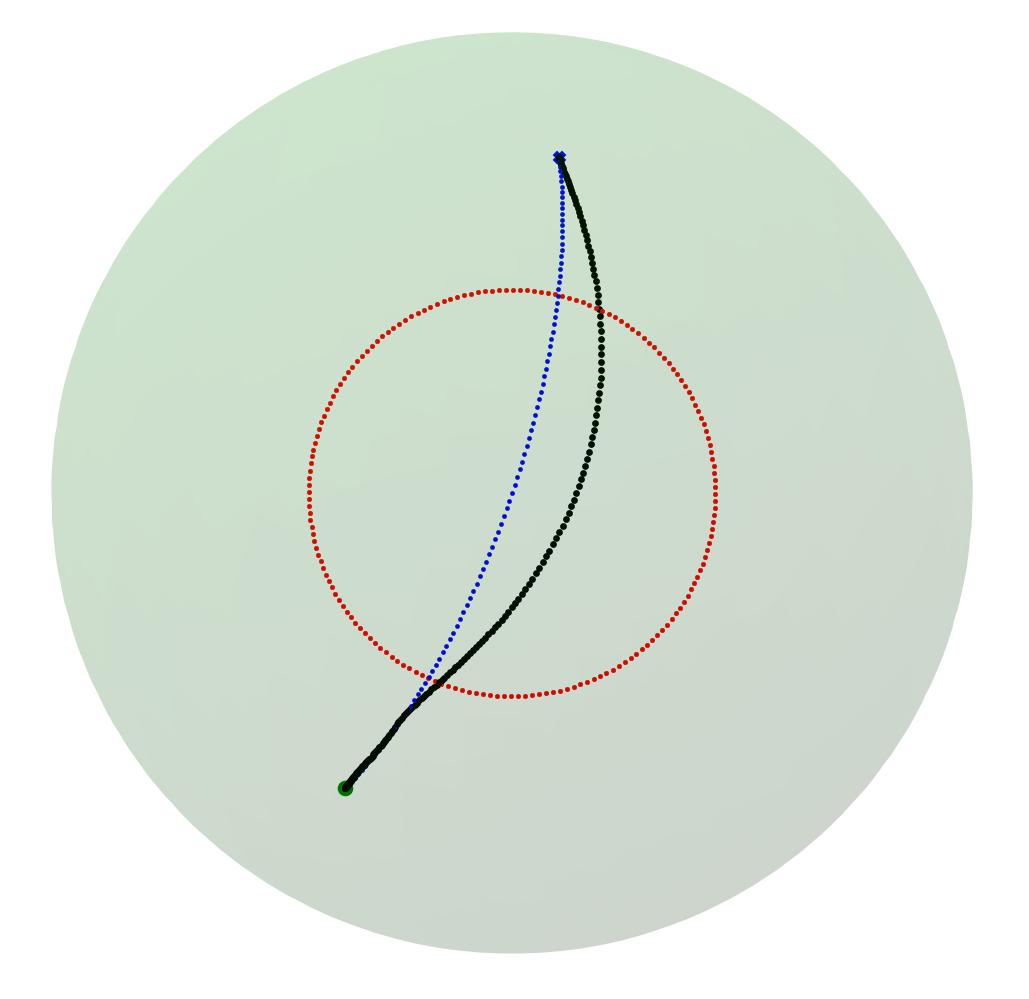}
\caption{Trace of boresight vector on unit-sphere with agent trained in Phase 1 (without F-zone).}
\label{fig:plot_3d_noFzone}
\end{figure}

\begin{figure}[h]
\centering
\includegraphics[width=.5\textwidth]
{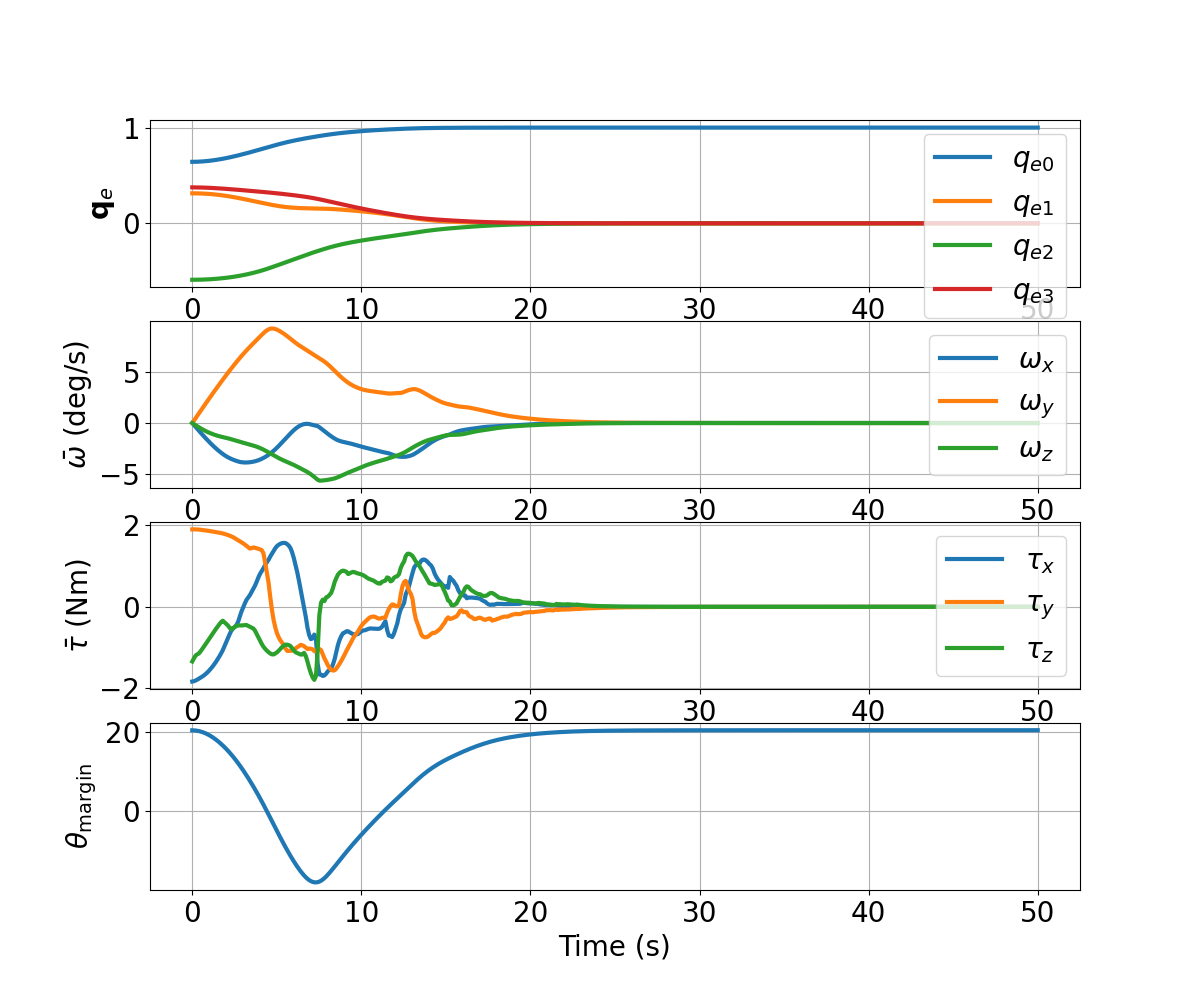}
\caption{Time history of relative attitude, angular velocity, control torque, and $\theta_\text{margin}$ under agent trained in Phase 1 (without F-zone).}
\label{fig:plot_data_noFzone}
\end{figure}

Figures~\ref{fig:plot_3d_noFzone} -~\ref{fig:plot_data_noFzone} present the results under the agent trained in Phase 1. Figure~\ref{fig:plot_3d_noFzone} visualizes the trace of the boresight vector (illustrated in black) on a unit sphere when controlled under the agent trained in Phase 1. The start and the end points are indicated as a blue cross and a green point, respectively. The red circle is the boundary of the F-zone. The boresight vector is required to not enter the area surrounded by the circle. It shows clearly that the boresight vector enters the F-zone during the maneuver controlled by the agent trained in Phase 1. The blue line indicates the trace of boresight vector when the spacecraft rotates along the shortest rotation path. Figure~\ref{fig:plot_data_noFzone} presents the time history of the relative attitude, the angular rate, the control torque, and $\theta_\text{margin}$. As the boresight vector enters the F-zone, $\theta_\text{margin}$ becomes negative.

\begin{figure}[h]
\centering
\includegraphics[width=.4\textwidth]
{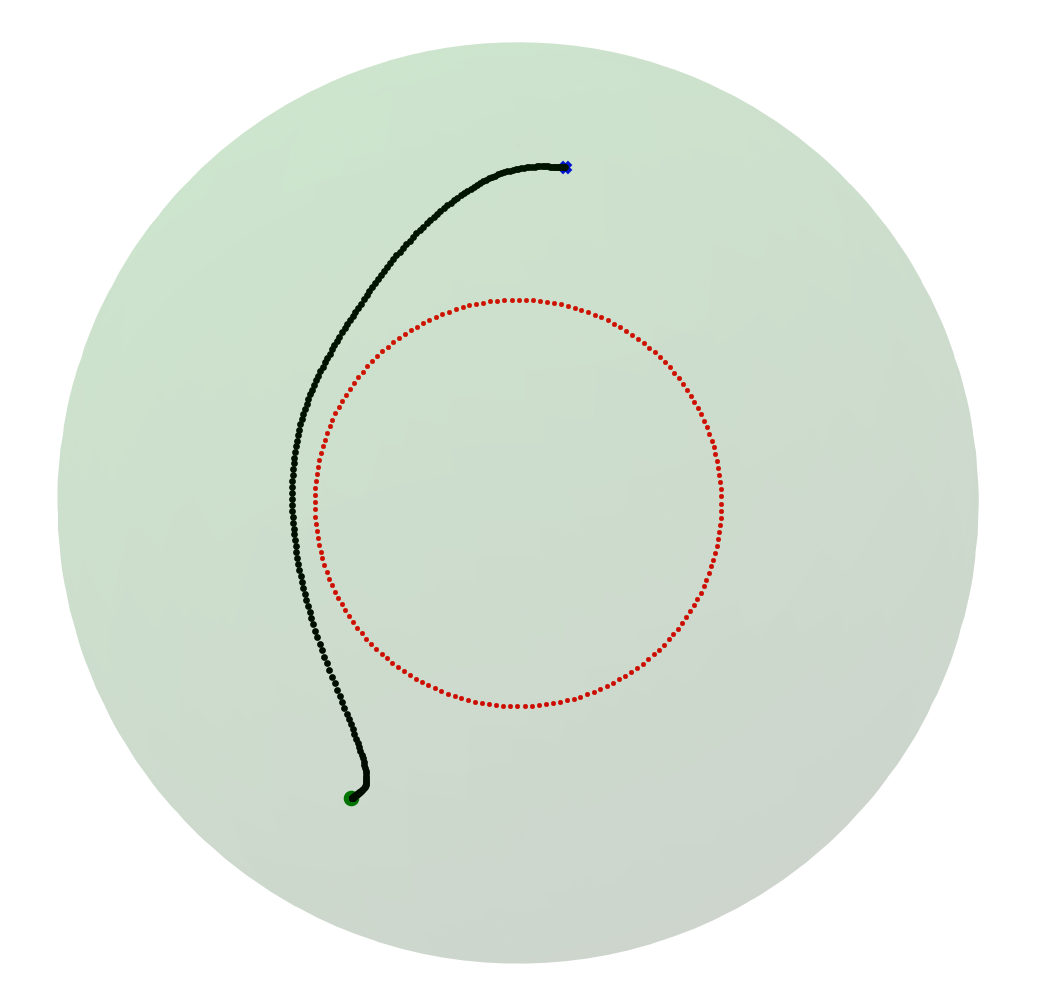}
\caption{Trace of boresight vector on unit-sphere with agent trained in Phase 2 (with one F-zone).}
\label{fig:plot_3d_withFzone}
\end{figure}

\begin{figure}[h]
\centering
\includegraphics[width=.5\textwidth]
{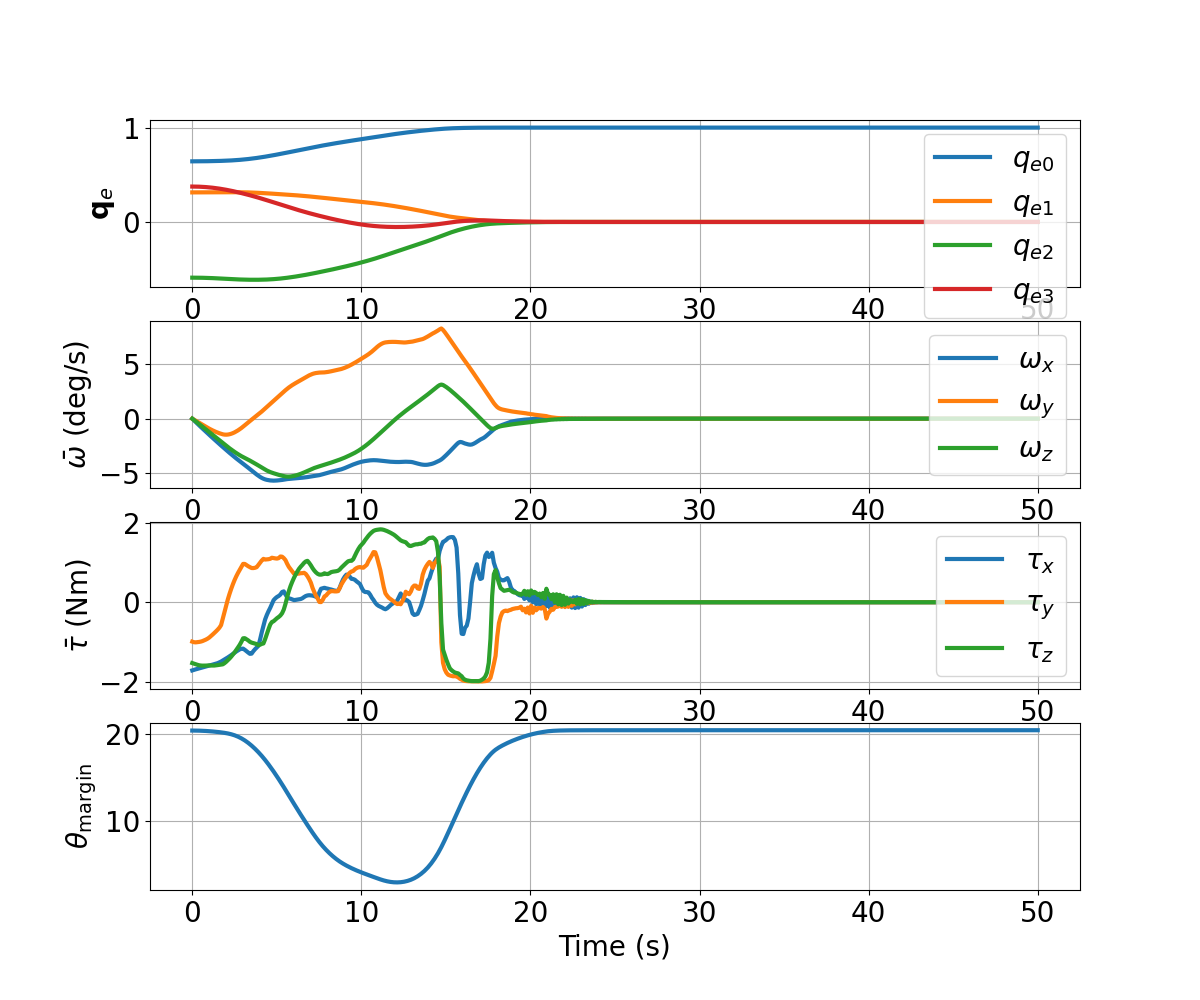}
\caption{Time history of relative attitude, angular velocity, control torque, and $\theta_\text{margin}$ under agent trained in Phase 2 (with one F-zone).}
\label{fig:plot_data_withFzone}
\end{figure}

Figures~\ref{fig:plot_3d_withFzone} -~\ref{fig:plot_data_withFzone} present results under the agent trained in Phase 2. Figure~\ref{fig:plot_3d_withFzone} illustrates the trace of the boresight vector (in black) on a unit sphere when controlled by the agent trained in Phase 2. The start and the end points of the trace and the F-zone are indicated in the same way as in Fig.~\ref{fig:plot_3d_noFzone}. The boresight vector moves to the desired pointing while avoiding the F-zone during the maneuver. Figure~\ref{fig:plot_data_withFzone} plots the time history of relevant data. As $\bm{q}_e$ reaches the identity quaternion, the desired pointing is achieved. $\theta_{\text{margin}}$ remains positive during the maneuver as the pointing keep-out zone is complied with. 

Results in Figs.~\ref{fig:plot_3d_withFzone} -~\ref{fig:plot_data_withFzone} demonstrate the effectiveness of the proposed design of state space and reward function for controlling the reorientation of spacecraft with a single pointing keep-out zone based on reinforcement learning.

\begin{figure}[h]
\centering
\includegraphics[width=.5\textwidth]
{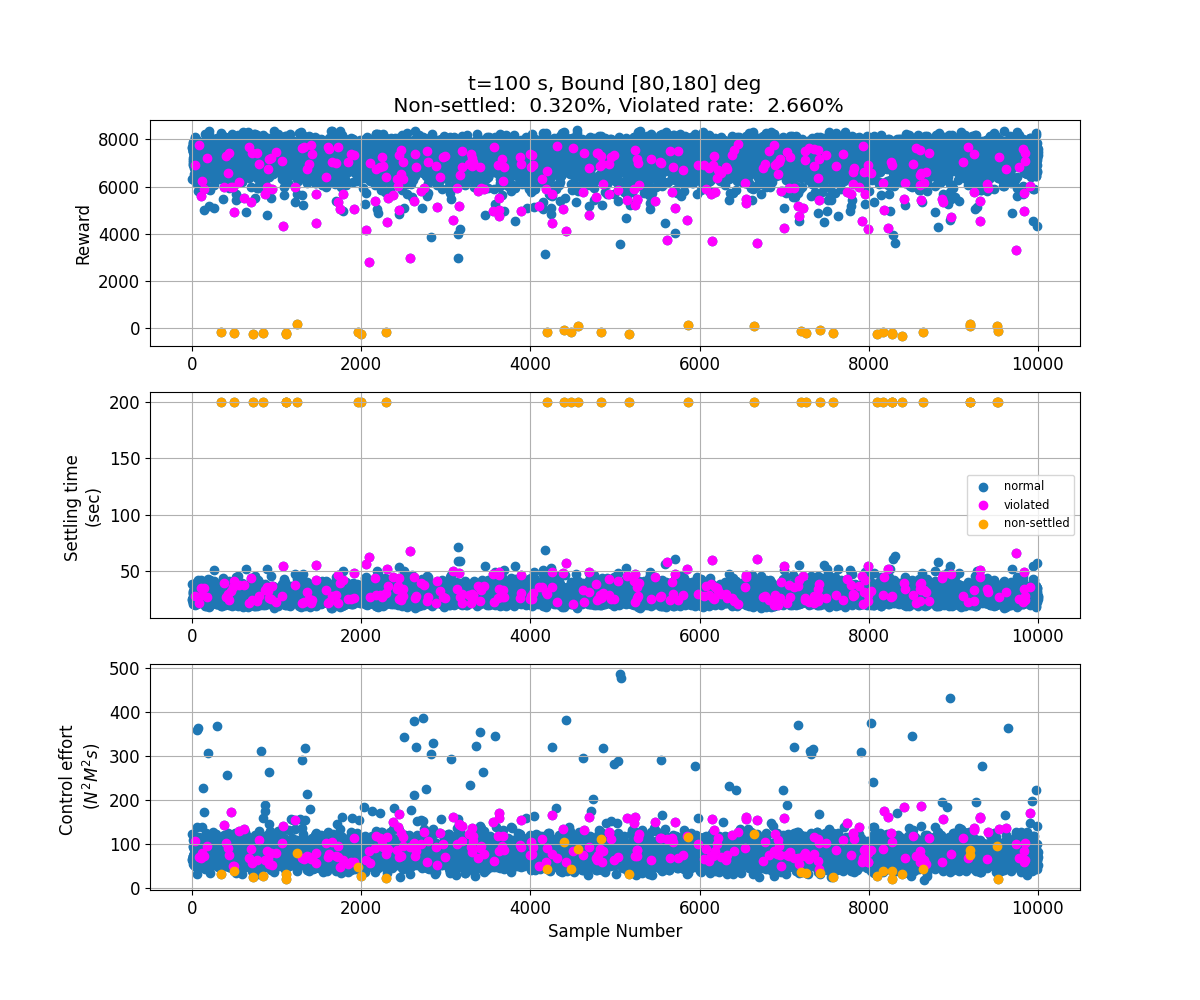}
\caption{Monte Carlo simulation results (metrics vs sample number) under the bested rewarded agent trained in Phase 2 (with one F-zone). Settling times for those non-settled cases are given a value of 200 s for the purpose of plotting.}
\label{fig:plot_monte_carlo_1}
\end{figure}

\begin{figure}[h]
\centering
\includegraphics[width=.5\textwidth]
{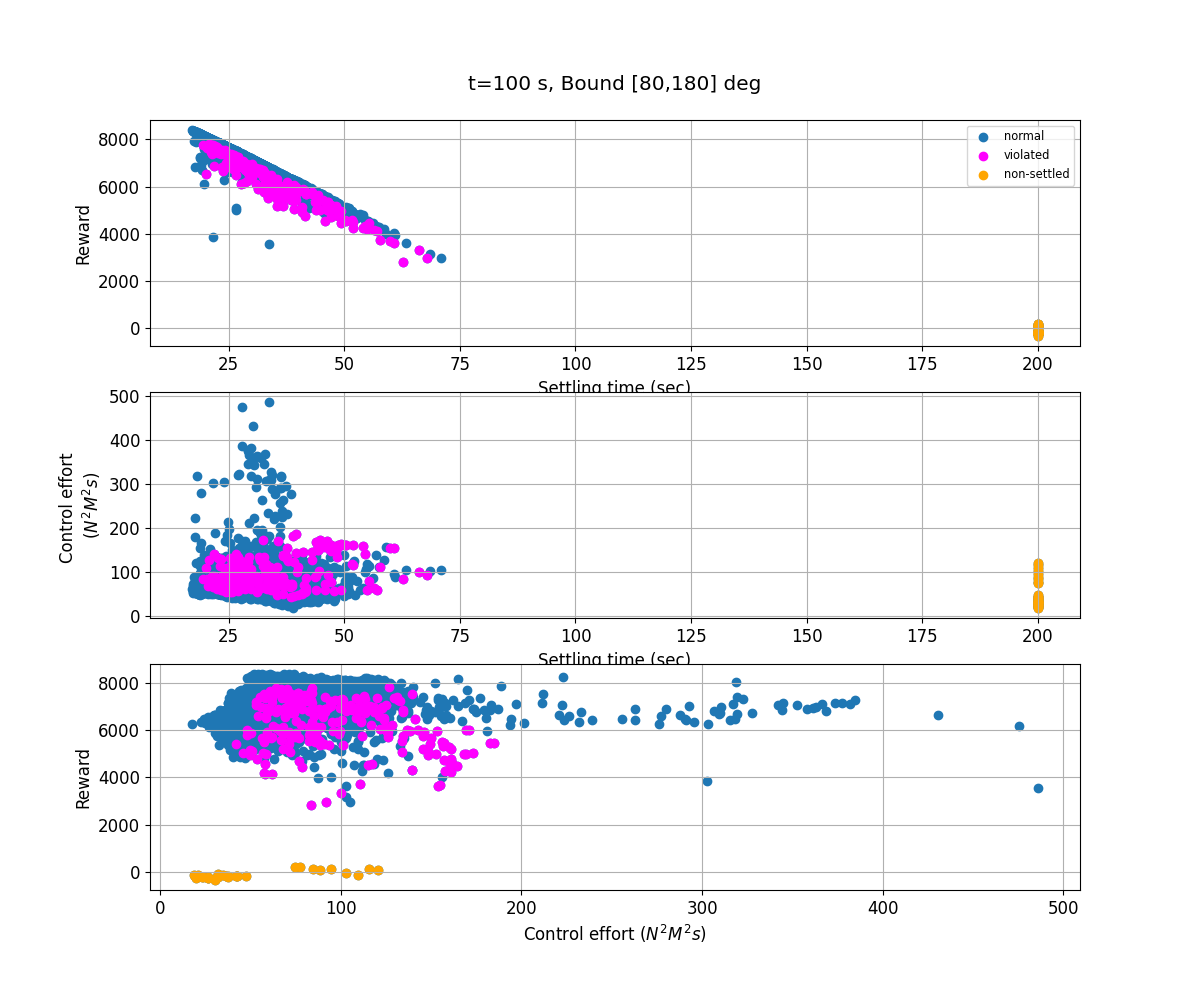}
\caption{Monte Carlo simulation results (relation between metrics) under the bested rewarded agent trained in Phase 2 (with one F-zone). Settling times for those non-settled cases are given a value of 200 s for the purpose of plotting.}
\label{fig:plot_monte_carlo_2}
\end{figure}

\begin{figure}[h]
\centering
\includegraphics[width=.5\textwidth]
{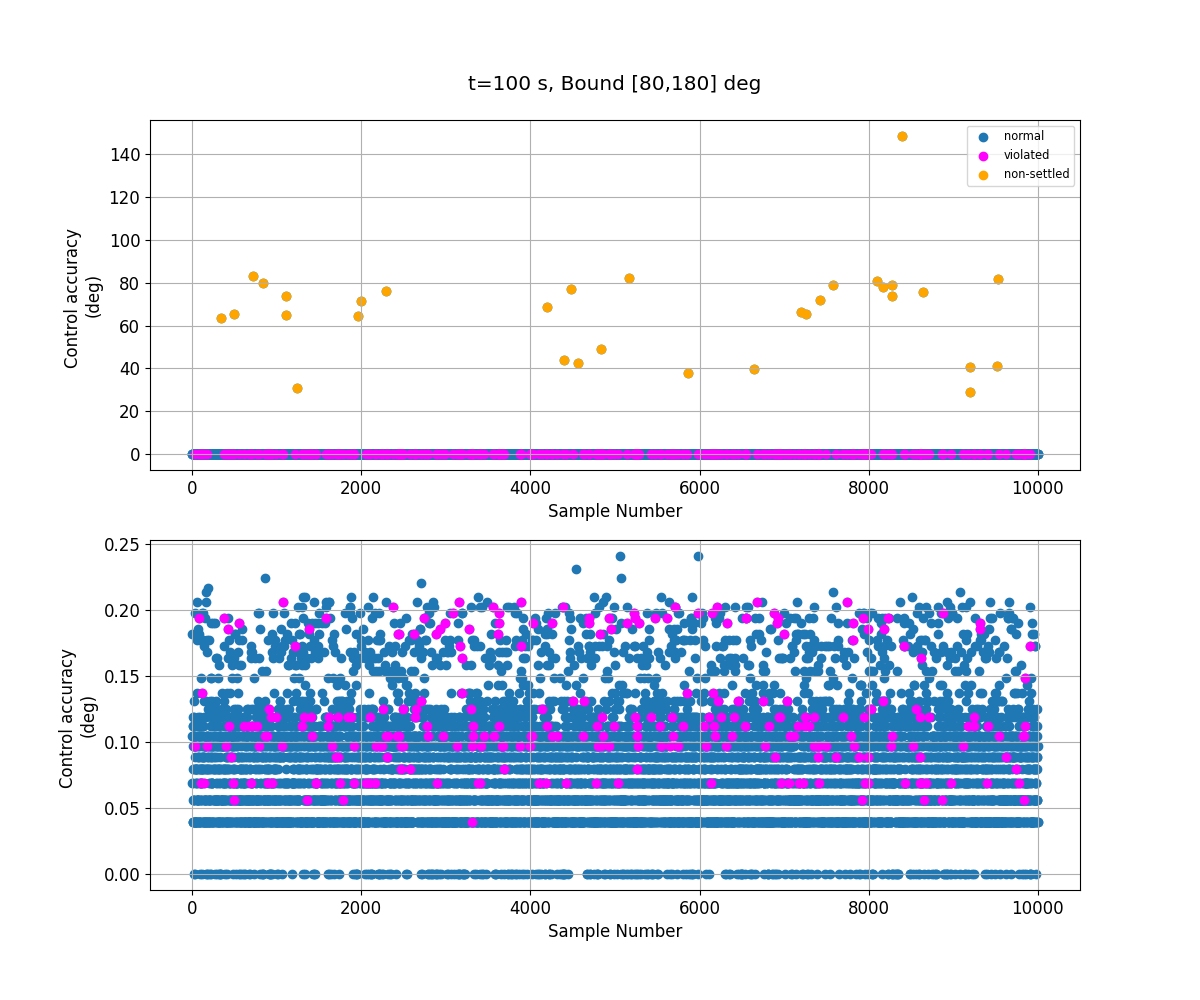}
\caption{Control accuracy under the bested rewarded agent trained in Phase 2 (with one F-zone). Top: non-settled cases included. Bottom: settled cases only.}
\label{fig:plot_monte_carlo_3}
\end{figure}

A Monte Carlo simulation is further performed to evaluate the performance of the best rewarded agent trained in Phase 2. 10000 scenarios with different combinations of initial deviation angles (random in the range $[80^{\circ},180^{\circ}]$) and initial angular rates (uniformly distributed in the range $[-0.001, 0.001]~^{\circ}$/s) are tested. The time step is set as~\SI{0.1}{s} and the duration per simulation is set as~\SI{100}{s}. The same moment of inertia as in the agent training is used for tests. For each test, the agent is evaluated based on the reward, the settling time, the control accuracy and the control effort which is calculated as \[E\left(t_{end}\right) = \int_{0}^{t_{end}} \|\bar{\tau}\|^2 dt\] The results are shown in Figs.~\ref{fig:plot_monte_carlo_1} -~\ref{fig:plot_monte_carlo_3} and Table~\ref{tab:monte_carlo_statistics}.\\
Figure~\ref{fig:plot_monte_carlo_1} presents rewards, settling times, and control efforts of the 10000 tests. 
In about $97\%$ of the simulations (samples in blue), the spacecraft can reach the desired orientation while complying with the constraint zone. The failed cases (about $3\%$) consist of those violating F-zone constraints (samples in purple, $2.66\%$) and those not settled within the simulation duration of 100 s (samples in orange, $0.32\%$). Note that, in Figs.~\ref{fig:plot_monte_carlo_1} -~\ref{fig:plot_monte_carlo_2}, for the purpose of plotting, settling times for those non-settled cases are given a value of 200 s.
A review of the non-settled cases shows that the attitude becomes stuck near its initial value or during the rotation toward the desired attitude. For the non-settled cases, very low rewards and low control efforts are observed. 

Figure~\ref{fig:plot_monte_carlo_2} shows the relation between the settling time, the reward, and the control effort for each test. As expected, the reward is closely related to the settling time. In general, cases with smaller settling times achieve higher rewards since an extra 9 point is rewarded when the desired accuracy is achieved as indicated in Eq.~(\ref{eq:reward_r}). There is no obvious relationship between the control effort and the settling time. The same applies to the relationship between the control effort and the reward.

Figure~\ref{fig:plot_monte_carlo_3} presents the control accuracy at the end of the simulation under the best agent for each test. For all settled cases, the desired accuracy ($0.25^\circ$ in this paper) is achieved at the end of the simulation.

The most important insight from the Monte Carlos simulation is that the reward shaping alone with consideration of F-zone cannot guarantee the compliance of attitude constraint during the maneuver. Further measures should be taken for the agent training in future work. One candidate method is safe RL, such as shielded RL, in which a shield mechanism is introduced to replace unsafe actions generated by the agent with safe ones.

\begin{table}[htbp]
\caption{\label{tab:monte_carlo_statistics} Results of Monte Carlo simulation (10000 tests)}
\centering
\begin{tabular}{cc}
\hline
\hline
Evaluation metrics & Value \\
\hline
Mean Reward  & $7281.91 \pm 688.85$ \\ 
Mean settling time* (sec) & $27.81 \pm 5.24 $ \\
Mean control effort* ($N^2m^2s$) & $76.02 \pm 25.76$ \\
Mean control accuracy* (deg) & $0.08 \pm 0.04$ \\
Rate of non-settled & $0.32\%$\\ 
Rate of violation & $2.66\%$ \\ 
\hline
\hline
* Non-settled cases are ignored.
\end{tabular}
\end{table}

\section{Conclusion}
This paper presents a DRL approach for spacecraft reorientation control with a single pointing keep-out zone. A novel state space representation and a tailored reward function are designed to address this problem. The SAC algorithm and a curriculum learning approach are used for agent training. 

Preliminary results demonstrate the effectiveness of the proposed design of state space and reward function in applying DRL to spacecraft reorientation control with a single pointing keep-out zone. The Monte Carlo simulation shows that the reward shaping alone cannot guarantee constraint-safe reorientation of spacecraft. Safe RL, such as shielded RL, needs to be implemented in the future work. Furthermore, future work will also consider cases involving multiple constraint zones for a boresight vector.

\begin{ack}
The authors would like to thank the JMU Seed Grant for its financial support.
\end{ack}

\bibliography{ifacconf}                         

@article{sutton2018reinforcement,
  title={Reinforcement learning: An introduction 2nd ed},
  author={Sutton, Richard S and Barto, Andrew G and others},
  journal={MIT press Cambridge},
  volume={1},
  number={2},
  pages={25},
  year={2018}
}

@article{arulkumaran2017deep,
  title={Deep reinforcement learning: A brief survey},
  author={Arulkumaran, Kai and Deisenroth, Marc Peter and Brundage, Miles and Bharath, Anil Anthony},
  journal={IEEE signal processing magazine},
  volume={34},
  number={6},
  pages={26--38},
  year={2017},
  publisher={IEEE}
}

@inproceedings{Djebko2023learning,
  title={Learning attitude control},
  author={Djebko, K. and Puppe, F. and Montenegro, S. and Baumann, T. and Faisal, M.},
  booktitle={14th IAA
Symposium on Small Satellites for Earth System Observation},
  year={2023}
}

@inbook{markley2014fundamentals,
  title={Fundamentals of Spacecraft Attitude Determination and Control},
  author={Markley, F Landis and Crassidis, John L},
  year={2014},
  publisher={Springer},
  address={New York},
  chapter={2,~3,~7}
}

@inproceedings{haarnoja2018soft,
  title={Soft actor-critic: Off-policy maximum entropy deep reinforcement learning with a stochastic actor},
  author={Haarnoja, Tuomas and Zhou, Aurick and Abbeel, Pieter and Levine, Sergey},
  booktitle={International conference on machine learning},
  pages={1861--1870},
  year={2018},
  organization={Pmlr}
}

@article{elkins2022bridging,
  title={Bridging reinforcement learning and online learning for spacecraft attitude control},
  author={Elkins, Jacob G and Sood, Rohan and Rumpf, Clemens},
  journal={Journal of Aerospace Information Systems},
  volume={19},
  number={1},
  pages={62--69},
  year={2022},
  publisher={American Institute of Aeronautics and Astronautics}
}

@inproceedings{gao2020satellite,
  title={Satellite attitude control with deep reinforcement learning},
  author={Gao, Duozhi and Zhang, Haibo and Li, Chuanjiang and Gao, Xinzhou},
  booktitle={2020 Chinese Automation Congress (CAC)},
  pages={4095--4101},
  year={2020},
  organization={IEEE}
}

@article{oghim2025deep,
  title={Deep reinforcement learning-based attitude control for spacecraft using control moment gyros},
  author={Oghim, Snyoll and Park, Junwoo and Bang, Hyochoong and Leeghim, Henzeh},
  journal={Advances in Space Research},
  volume={75},
  number={1},
  pages={1129--1144},
  year={2025},
  publisher={Elsevier}
}

@article{cai2024reinforcement,
  title={Reinforcement learning-based satellite formation attitude control under multi-constraint},
  author={Cai, Yingkai and Low, Kay-Soon and Wang, Zhaokui},
  journal={Advances in Space Research},
  volume={74},
  number={11},
  pages={5819--5836},
  year={2024},
  publisher={Elsevier}
}

@inproceedings{jiang2023spacecraft,
  title={Spacecraft Attitude Maneuver Planning Based on Deep Reinforcement Learning under Complex Constraints},
  author={Jiang, Shulei and Zhao, Fanyu and Chen, Yuejie and Jin, Zhonghe},
  booktitle={2023 9th International Conference on Control Science and Systems Engineering (ICCSSE)},
  pages={61--67},
  year={2023},
  organization={IEEE}
}

@inproceedings{feron2001randomized,
  title={A randomized attitude slew planning algorithm for autonomous spacecraft},
  author={Feron, E and Dahleh, M and Frazzoli, E and Kornfeld, R},
  booktitle={AIAA guidance, navigation, and control conference and exhibit},
  pages={4155},
  year={2001}
}

@article{kjellberg2013discretized,
  title={Discretized constrained attitude pathfinding and control for satellites},
  author={Kjellberg, Henri C and Lightsey, E Glenn},
  journal={Journal of Guidance, Control, and Dynamics},
  volume={36},
  number={5},
  pages={1301--1309},
  year={2013},
  publisher={American Institute of Aeronautics and Astronautics}
}

@article{lee2014feedback,
  title={Feedback control for spacecraft reorientation under attitude constraints via convex potentials},
  author={Lee, Unsik and Mesbahi, Mehran},
  journal={IEEE Transactions on Aerospace and Electronic Systems},
  volume={50},
  number={4},
  pages={2578--2592},
  year={2014},
  publisher={IEEE}
}

@article{yang2021potential,
  title={Potential field-based sliding surface design and its application in spacecraft constrained reorientation},
  author={Yang, Juntang and Duan, Yisheng and Ben-Larbi, Mohamed Khalil and Stoll, Enrico},
  journal={Journal of Guidance, Control, and Dynamics},
  volume={44},
  number={2},
  pages={399--409},
  year={2021},
  publisher={American Institute of Aeronautics and Astronautics}
}

@article{raffin2021stable,
  title={Stable-baselines3: Reliable reinforcement learning implementations},
  author={Raffin, Antonin and Hill, Ashley and Gleave, Adam and Kanervisto, Anssi and Ernestus, Maximilian and Dormann, Noah},
  journal={Journal of machine learning research},
  volume={22},
  number={268},
  pages={1--8},
  year={2021}
}
\end{document}